\documentclass[10pt, conf]{IEEEtran}
\ifCLASSINFOpdf
   \usepackage[pdftex]{graphicx}
   \graphicspath{{../figures/}{../plots/}}
\else
   \usepackage[dvips]{graphicx}
   \graphicspath{{../figures/}{../plots/}}
\fi
\usepackage[cmex10]{amsmath}
\usepackage{amssymb,verbatim}
\usepackage{amsthm,bm}
\usepackage{algorithmic}
\usepackage{algorithm}
\usepackage[tight,footnotesize]{subfigure}
\newcommand{\B}{\mathcal{B}}
\newcommand{\C}{\mathcal{C}}
\newcommand{\F}{\mathcal{F}}
\newcommand{\myprob}[1]{\mathsf{P}\left\{#1\right\}}
\newcommand{\myexp}[1]{\mathsf{E}\left[#1\right]}
\newtheorem{lemma}{Lemma}

\newcommand{\nn}{\nonumber\\}

\usepackage{color}

\begin{document}
\title{Utility Optimal Coding for Packet Transmission over Wireless
       Networks -- Part I:\\ Networks of Binary Symmetric Channels}
\author{K.~Premkumar, Xiaomin Chen, and Douglas J.~Leith\\
        Hamilton Institute, National University of Ireland, Maynooth,
		Ireland\\
		E--mail: \{Premkumar.Karumbu, Xiaomin.Chen, Doug.Leith\}@nuim.ie
		 
        \thanks{This work is supported by Science Foundation Ireland
		under Grant No. 07/IN.1/I901.}
       }
\maketitle
\thispagestyle{empty}
\pagestyle{empty}
\begin{abstract}
We consider multi--hop networks comprising Binary Symmetric Channels
($\mathsf{BSC}$s). The network carries unicast flows for multiple users.
The utility of the network is the sum of the utilities of the flows,
where the utility of each flow is a concave function of its throughput.
Given that the network capacity is shared by the flows, there is a
contention for network resources like coding rate (at the physical
layer), scheduling time (at the MAC layer), etc., among the flows. We
propose a proportional fair transmission scheme that maximises the sum
utility of flow throughputs subject to the rate and the scheduling
constraints. This is achieved by {\em jointly optimising the packet
coding rates of all the flows through the network}.
\end{abstract}


%
\IEEEpeerreviewmaketitle

\section{Introduction}
In a communication network, the network capacity is shared by a set of
flows. There is a contention for resources among the flows, which leads
to many interesting problems. One such problem, is {\em how to allocate
the resources optimally across the (competing) flows, when the physical
layer is erroneous}. Specifically, schedule/transmit time for a flow is
a resource that has to be optimally allocated among the competing flows.
In this work, we pose a network utility maximisation problem subject to 
scheduling constraints that solve a resource allocation problem.

\begin{figure}[t]
\centering
\includegraphics[width=75mm, height=75mm]{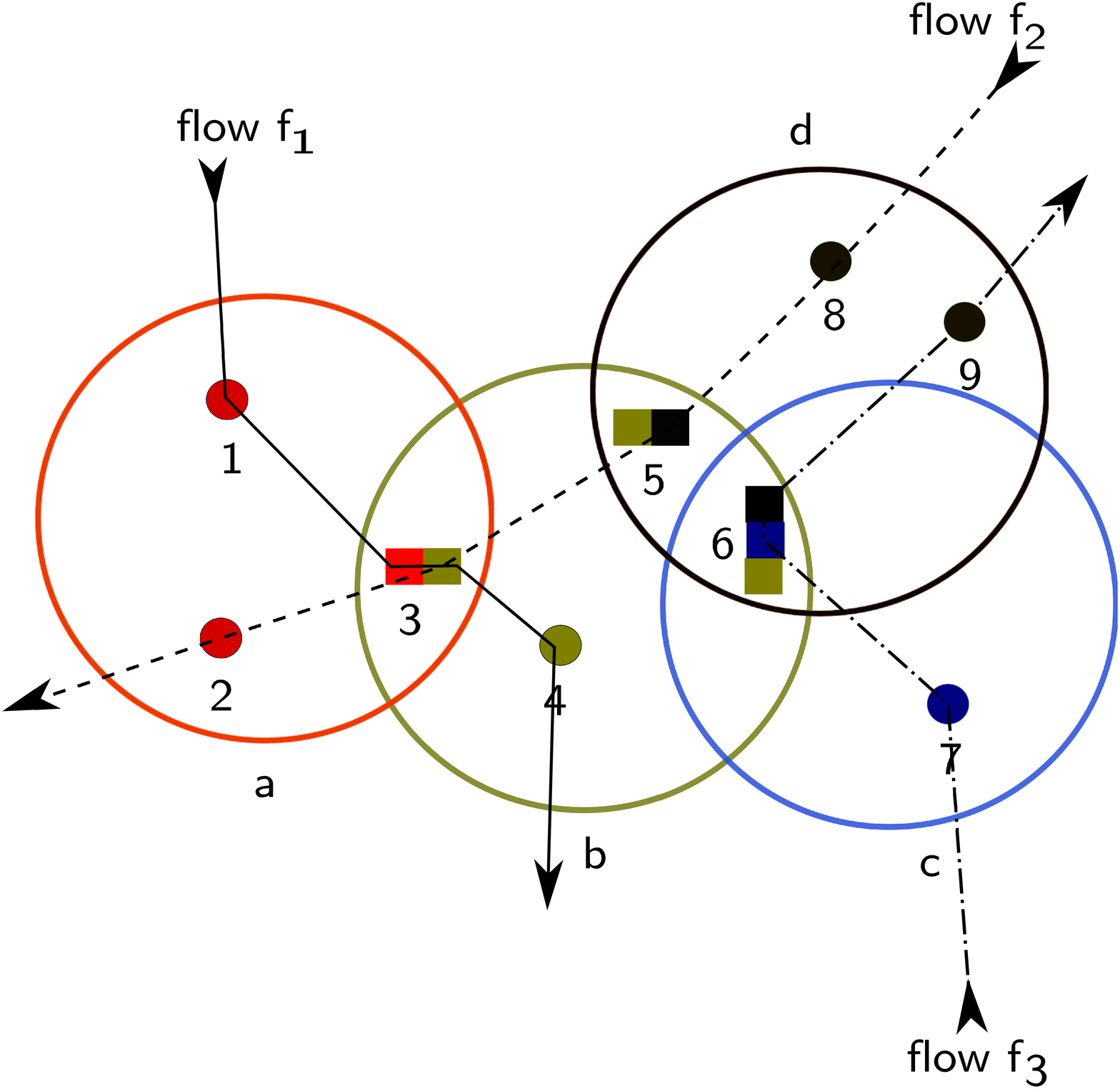}
\caption{{\bf An illustration of a wireless mesh network with 4 cells.}
Cells $a$, $b$, $c$, and $d$ use orthogonal channels CH$_1$, CH$_2$,
CH$_3$, and CH$_4$ respectively. Nodes 3, 5, and 6 are {\em bridge
nodes}. The bridge node 3 (resp. 5 and 6) is provided a time slice of
each of the channels CH$_1$ \& CH$_2$ (resp. CH$_2$ \& CH$_4$ for node 5
and CH$_2$\& CH$_3$\& CH$_4$ for node 6). Three flows $f_1, f_2$, and
$f_3$ are considered. In this example, $\C_{f_1} = \{a,b\}$, $\C_{f_2} =
\{d,b,a\}$, and $\C_{f_3} = \{c,d\}$.} 
\label{fig:mesh_network}
\end{figure}

We consider packet communication over multi--hop networks comprising of
Binary Symmetric Channels ($\mathsf{BSC}$s, \cite{cover_book}). The
network consists of a set of $C \geq 1$ cells $\C = \{1,2,\cdots,C\}$
which define the ``interference domains'' in the network. We allow
intra--cell interference (\emph{i.e} transmissions by nodes within the
same cell interfere) but assume that there is no inter--cell
interference. This captures, for example, common network architectures
where nodes within a given cell use the same radio channel while
neighbouring cells using orthogonal radio channels. Within each cell,
any two nodes are within the decoding range of each other, and hence,
can communicate with each other. The cells are interconnected using
multi--radio bridging nodes to create a multi--hop wireless network. A
multi--radio bridging node $i$ connecting the set of cells
$\B(i)=\{c_1,..,c_n\}\subset \C$ can be thought of as a set of $n$
single radio nodes, one in each cell, interconnected by a high--speed,
loss--free wired backplane (see Figure~\ref{fig:mesh_network}).

Data is transmitted across this multi--hop network as a set $\F$ $=$
$\{1,2,\cdots,F\}$, $F\geq 1$ of unicast flows. The route of each flow
$f$ $\in \F$ is given by $\C_f$ $=$ $\{c_1(f), c_2(f), \cdots,
c_{\ell_f}(f)\}$, where the source node $s(f) \in c_1(f)$ and the
destination node $d(f) \in c_{\ell_f}(f)$. We assume loop--free flows
(i.e., no two cells in $\C_f$ are same).  Figure~\ref{fig:mesh_network}
illustrates this network setup. A scheduler assigns a time slice of
duration $T_{f,c} > 0$ time units to each flow $f$ that flows through
cell $c$, subject to the constraint that $\sum_{f:c\in \C_f}
T_{f,c}\le T_c$ where $T_c$ is the period of the schedule in cell $c$.
We consider a periodic scheduling strategy in which, in each cell $c$,
service is given to the flows in a round robin fashion, and that each
flow $f$ in cell $c$ gets a time slice of $T_{f,c}$ units in every schedule.

The scheduled transmit times for flow $f$ in source cell $c_1(f)$ define
time slots for flow $f$. We assume that a new information packet arrives
in each time slot, which allows us to simplify the analysis by ignoring
queueing. Information packets of each flow $f$ at the source node $S(f)$
consist of a block of $k_f$ symbols. Each packet of flow $f$ is encoded
into codewords of length $n_f = k_f/r_f$ symbols, with coding rate
$0< {r_f}\le1$. The code employed for encoding is discussed in Section 
\ref{sec:problem_formulation}. We require sufficient transmit times 
at each cell along route $\C_f$ to allow $n_f$ coded symbols to be 
transmitted in every schedule period. Hence there is no queueing at the 
cells along the route of a flow.

{\bf Channel Model:}
The channel in cell $c$ for flow $f$ is considered to be a binary
symmetric channel ($\mathsf{BSC}$) with the cross--over probability
(i.e., the probability of a bit error) being $\alpha_{f,c} \in
[0,1]$. The corresponding transition probability matrix is thus given by
\begin{eqnarray*}
{\bf H}_{f,c}(\alpha_{f,c}) & = & \left[ 
\begin{array}{cc}
1-\alpha_{f,c} &  \alpha_{f,c} \\ 
  \alpha_{f,c} & 1-\alpha_{f,c} 
\end{array}
\right].
\end{eqnarray*}
Thus, the end--to--end channel for flow $f$ is a cascaded channel (of
$\ell_f$ $\mathsf{BSC}$s), which is a $\mathsf{BSC}$, with the
transition probability matrix ${\bf H}_f(\alpha_f) = \prod_{c \in \C_f} {\bf
H}_{f,c}(\alpha_{f,c})$, the cross--over probability of which is given by 
\begin{eqnarray*}
\alpha_f = \sum_{\{ x_c \in \{0,1\}, c\in\C_f:\underset{c \in
\C_f}{\sum} x_c \ \text{is odd}\}} \ \ 
\prod_{c \in \C_f} \alpha_{f,c}^{x_c}
\ \left(1-\alpha_{f,c}\right)^{1-x_c}.
\end{eqnarray*}
Since, each transmitted symbol in a packet of a flow can, in general,
take values from a $2^m = M$--ary alphabet, there are $m$ channel uses
of the $\mathsf{BSC}$ for every transmitted symbol. Thus, the symbol
error probability (for any $m \geq 1$) is given by $\beta_f = 1 -
(1-\alpha_f)^{m}$. Let the Bernoulli random variable $E_{f}[i]$
indicate the end--to--end error of the $i$th coded symbol at the
destination in a code word of flow $f$. Note that $E_f[i]$s are 
independent and identically distributed (i.i.d.), and that 
${\sf P}\{E_f[i]=1\}= \beta_f = 1 - {\sf P}\{E_f[i]=0\}$. In the
channel model described, the channel processes across time are
independent copies of the $\mathsf{BSC}$s. This is realised in a
wireless network by means of an interleaver of sufficient depth (after
the channel encoder), which interleaves the encoded symbols. The interleaved symbols see a fading
channel (which is modelled as a channel with memory, e.g., a
Gilbert--Elliot channel \cite{mushkin1989capacity}), but  the
de--interleaver (before the channel decoder) brings back the
original sequence of the encoded symbols, but interleaves the channel
fades, the combined effect of which can be modelled as independent
channel processes across time. In another work \cite{erasure}, we model the
fading channel as a packet erasure channel (or a block fading channel), and obtain the optimal
transmission strategy, which includes optimal interleaving of bits
across schedules and the optimal coding rates.

Letting $e_f(r_f)$ denote the error probability that a packet fails to
be decoded, the expected number of information symbols successfully
received is $S_f(r_f)=k_f(1-e_f(r_f))$. Other things being equal, one
expects that decreasing $r_f$ (i.e., increasing the number of redundant
symbols $n_f - k_f$) decreases error probability $e_f$, and so increases
$S_f$. However, since the network capacity is limited, and is shared by
multiple flows, increasing the coded packet size $n_{f_1}$ of flow $f_1$
generally requires decreasing the packet size $n_{f_2}$ for some other
flow $f_2$. That is, increasing $S_{f_1}$ comes at the cost of
decreasing $S_{f_2}$. {\em We are interested in understanding this
trade--off, and in analysing the optimal fair allocation of coding rates
amongst users/flows}.

\vspace{3mm}

\noindent
{\bf Contributions:}
Our main contribution is the analysis of fairness in the allocation of
coding rates between users/flows competing for limited network capacity.
In particular, we pose a resource allocation problem in the
utility--fair framework, and propose a scheme for obtaining the
proportional fair allocation of coding rates, \emph{i.e.} the allocation
of coding rates that maximises $\sum_{f\in\F}\log S_f(r_f)$ subject to
network capacity constraints (or scheduling constraints).  Specifically,
at the physical layer, the (channel) coding rate of a flow can be
lowered (to alleviate its channel errors) only at the expense of
increasing the coding rates of other flows. Also, at the network layer,
the length of schedules of each flow should be chosen in such a way that
it maximises the network utility. Interestingly, we show in our problem
formulation that the coding rate and the scheduling are tightly coupled.
Also, we show that for a $\log$ (network) utility function (which typically gives 
proportional fair allocation of resources) the optimum rate allocation
(in general) gives unequal air--times which is quite different from the previously
known result of proportional fair allocation being the same as that of
equal air--time allocation (\cite{eq_air_time}).
This problem, which we show in Section~\ref{sec:num}, requires solving a
non--convex optimisation problem. Our work differs from the previous
work on network utility maximisation (see \cite{net-opt} and the
references therein) in the following manner. To the best of our
knowledge, this is the first work that computes the optimal coding rate
for a given scheduling (or capacity) constraints in the utility--optimal
framework.

The rest of the paper is organised as follows. In
Section~\ref{sec:problem_formulation}, we obtain a measure for the
end--to--end packet decoding error, and describe the throughput of the
network. In Section~\ref{sec:num}, we formulate a network utility
maximisation problem subject to constraints on the transmission schedule
lengths. We obtain the optimum coding rates for each flow in the network
in Section~\ref{sec:sol}. In Section~\ref{sec:examples}, we provide some
simple examples to illustrate our results. The proofs of various Lemmas
are omitted due to lack of space.

\section{Packet Error Probability}
\label{sec:problem_formulation}

We recall that each transmitted symbol of flow $f$ reaches the
destination node erroneously with probability $\beta_f$. 
Hence, to recover the information
packets, we employ a block code at the source nodes (a convolutional
code with zero--padding is also a block code). 
Since an $(n,k,d)$ code can correct up to $\lfloor \frac{d-1}{2}
\rfloor$ errors, we are interested in employing a code with a large
distance $d$. Thus, a natural choice is the class of
(linear) maximum--distance separable (MDS) codes. MDS codes of rate $k/n$ have the
property that it achieves the Singleton bound (\cite{mc_williams_sloane}), 
\newline
\begin{minipage}{\columnwidth}
\begin{align}
d   & \leqslant n - k +1,
\end{align}
\normalsize
\end{minipage}
\newline
i.e., the minimum distance between any two codewords $d$, in an MDS code
is $n-k+1$. Thus, the maximum number of errors that an MDS code
can correct is $\left\lfloor\frac{d-1}{2}\right\rfloor =
\left\lfloor\frac{n-k}{2}\right\rfloor$. It is well known that in the
case of binary signalling, only trivial MDS codes exist. Hence, in this
paper, we consider $M =2^m$--ary alphabet, where $m > 1$. Examples for
MDS codes in the case of non--binary alphabets include Reed--Solomon
codes (\cite{mc_williams_sloane}), and  MDS--convolutional codes
(\cite{mds_conv_codes}). In \cite{mds_conv_codes}, the authors show the
existence of MDS--convolutional codes for any code rate. 
We note here that Reed--Solomon codes can also correct burst errors, and
hence, is more suitable for wireless networks (which does not employ an
interleaver).

\subsection{Network Constraints on Coding Rate}
Based on the modulation and the bandwidth available at each cell $c$, a
flow $f$, which passes through it, can obtain a maximum feasible physical
(PHY) rate of transmission in bits per second that the cell $c$ can
support.  Let $w_{f,c}$ be the PHY rate of transmission of flow $f$ in cell $c$.
For each transmitted packet of flow $f$, each cell $c \in \C_f$ along
its route must allocate at least $\frac{n_f}{w_{f,c}}$ units of time to
transmit the packet (or encoded block) where we recall that $n_f$ is the
length of the code word. Let $\F_c := \{f \in \F: c \in \C_f\}$ be the
set of flows that are routed through cell $c$. We recall that the
transmissions in any cell $c$ are scheduled in a TDMA fashion, and
hence, the total time required for transmitting packets for all flows in
cell $c$ is given by $\sum_{f \in \F_c} \frac{n_f}{w_{f,c}}$.  Since,
for cell $c$, the transmission schedule interval is $T_c$ units of time,
the coding rates $r_f$ must satisfy the schedulability constraint
$\sum_{f \in \F_c} \frac{k_f}{r_f w_{f,c}} \leqslant T_c$.

\subsection{Error Probability -- Upper bound}
The symbol errors $E_f[1], E_f[2], \cdots, E_f[n_f]$ are i.i.d.\
Bernoulli random variables, and hence, the probability of a codeword
(or encoded packet) being decoded incorrectly is given by ${\sf
P}\left\{\sum_{i=1}^{n_f}E_f[i] > \frac{n_f-k_f}{2} \right\}$.  We
observe that $\sum_{i=1}^{n_f}E_f[i]$ is a binomial random variable, and
hence, the probability of decoding error can be computed exactly.  However, the
exact probability of error is not tractable for further optimisation as
the probability of error, which is a function of the coding rate, is
neither concave nor convex. Hence, we pose the 
problem based on the upper bound on the error probability So, we obtain
an upper bound and a lower bound for the error probability. We show that
the bounds are tight, and hence, the problem of network utility
maximisation can be posed based on the lower bound on the error
probability. 

\begin{lemma}
\label{lem:pe_bound}
An upper bound for the end--to--end probability of a packet decoding error for flow $f$ is
bounded by the following.
\begin{align}
\label{eqn:pe_bound}
\widetilde{e}_f &= \
  \myprob{\sum_{i=1}^{n_f}E_f[i] > \frac{n_f-k_f}{2}} \nn
  &\leq \ \exp\left(-\frac{k_f}{r_f} I_{E_f[1]}\left(\frac{1-r_f}{2};\theta_f\right)\right)\\
  &=:   \ e_f(\theta_f,r_f).\nonumber
\end{align}
where $\theta_f >0$ is the Chernoff--bound parameter and the function
$I_Z(x;\theta) := \theta x - \ln(\myexp{e^{\theta Z}})$ is called the
rate function in large deviations theory.
\end{lemma}

\subsection{Error Probability -- Lower bound}

\begin{lemma}
\label{lem:pe_lower_bound}
The end--to--end probability of a packet decoding error for flow $f$ is
at least as large as 
\begin{align}
\widetilde{e}_f &\ge  
 \left[\frac{\beta_f}{1-\beta_f} 
 \exp\left(-\frac{k_f}{1-2x_f} H({\cal B}(x_f))\right) \right]\nn
 &\cdot \exp\left(-\frac{k_f}{1-2x_f} D({\cal B}(x_f)\|{\cal B}(\beta_f))\right)
\end{align}
where ${\cal B}(x)$ is the Bernoulli distribution with parameter $x$,
$H({\cal P})$ is the entropy of probability mass function (pmf) ${\cal
P}$, and $D({\cal P}\|{\cal Q})$ is the information divergence between the
pmfs ${\cal P}$ and ${\cal Q}$. 
\end{lemma}

From the lower and the upper bounds for the probability of packet
decoding error, and for the optimal $\theta_f^*$ (see
Eqn.~\eqref{e_f_star} in Section~\ref{sec:sol}), we see that the
exponent of the lower bound is the same as that of the upper bound
(Eqn.~\eqref{e_f_star}) with a pre--factor.  This motivates us to work
with the lower bound $e_f$ as a candidate to compute the utility of flow
$f$, which is given by $\ln(k_f(1-e_f))$.

We recall that $E_f[1]$ is a Bernoulli random variable which takes 
1 with probability $\beta_f$, and 0 with probability $1-\beta_f$. Thus 
$I_{E_f[1]}\left(\frac{1-r_f}{2};\theta_f\right) = \theta_f
\left(\frac{1-r_f}{2}\right) - \ln\left(1-\beta_f+\beta_f e^{\theta_f}\right)$.
Let $x_f := \frac{1-r_f}{2}$. Note that $0 \leqslant x_f < \frac{1}{2}$.
Therefore, from Eqn.~\eqref{eqn:pe_bound}, 
\begin{align}
\label{eqn:e_bound}
e_f(\theta_f,x_f) &:= \exp\left(-\frac{k_f}{1-2x_f}\left[
\theta_f x_f - \ln\left(1-\beta_f+\beta_f e^{\theta_f}\right)
\right]\right)
\end{align}

\vspace{4mm}

\section{Network Utility Maximisation} 
\label{sec:num}
We are interested in maximising the utility of the network which is
defined as the sum utility of flow throughputs. We consider the log of
throughput as the candidate for the utility function being motivated by
the desirable properties like proportional fairness that it possesses. 

We define the following notations: Chernoff--bound parameters 
${\bm \theta}:=[\theta_f]_{f \in \F}$, code rates ${\bm r} := [r_f]_{f
\in \F}$, and $x$ parameters ${\bm x} := [x_f]_{f \in \F}$ (where we
recall that $x_f = (1-r_f)/2$). We define the network utility as  
\begin{eqnarray}
\widetilde{U}\left({\bm \theta}, {\bm x}\right)
& := & \sum_{f \in \F} \ln\left(k_f \left(1 - e_f(\theta_f,x_f)\right)\right)\nn
&  = & \sum_{f \in \F} \ln\left(k_f\right) + 
       \sum_{f \in \F} \ln \left(1 - e_f(\theta_f,x_f)\right).
\end{eqnarray}
The problem is to obtain the optimum coding rate parameter ${\bm x}^*$
and the optimum Chernoff--bound parameter ${\bm \theta}^*$, which
maximises the network utility. Since, $k_f$, the size of information
packets of each flow $f$ is given, maximising the network utility is
equivalent to maximising 
\begin{eqnarray}
U({\bm \theta},{\bm x}) & := & \sum_{f\in \F} \ln\left(1-e_f(\theta_f,x_f)\right). 
\end{eqnarray}
Thus, we define the following
problem
\newpage
{\bf P1:}
\newline
\begin{minipage}{\columnwidth}
\small
\begin{align}
\max_{{\bm \theta},{\bm x}} \ \ \ & 
U({\bm \theta},{\bm x}) =  
\sum_{f \in \F} \ln\left(1-e_f(\theta_f,x_f)\right) \nn
\text{subject to} \quad
&\underset{f: c \in \C_f  }{\sum} \frac{k_f}{(1-2x_f)w_{f,c}}  \leq  T_c,  &&\forall c \in \C \label{eq:cons1} \\ 
 &\theta_f  >  0, &&\forall f \in \F                       \nn
 & x_f  \leq  \overline{\lambda}_f \,  &&\forall f \in \F  \nn
 & x_f  \geq  \underline{\lambda}_f \, &&\forall f \in \F  \nn
\end{align}
\normalsize
\end{minipage}
\newline
We note that the Eqn.~\eqref{eq:cons1} enforces the network capacity (or
the network schedulability) constraint. The objective function $U({\bm
\theta},{\bm x})$ is separable in $(\theta_f,x_f)$ pair for each flow
$f$. Importantly, the component of utility function for each flow $f$ given by
$\ln\left(1-e_f(\theta_f,x_f)\right)$ is not jointly concave in
$(\theta_f,x_f)$. However, $\ln\left(1-e_f(\theta_f,x_f)\right)$ is
concave in $\theta_f$ (for any $x_f$), and in $x_f$ (for any
$\theta_f$). Hence, the network utility maximisation problem ${\bf P1}$
is not in the standard convex optimisation framework. Instead, we pose
the following problem,
\newline

\noindent
{\bf P2:}
\newline
\begin{minipage}{\columnwidth}
\small
\begin{align}
\max_{\bm \theta} \max_{\bm x} 
& 
\sum_{f \in \F} \ln\left(1-e_f(\theta_f,x_f)\right) \label{eqn:coordinate_max_problem} \\
\text{subject to} \quad
&\underset{f: c \in \C_f  }{\sum} \frac{k_f}{(1-2x_f)w_{f,c}}  \leq  T_c,  &&\forall c \in \C \nonumber \\ 
 &\theta_f  >  0, &&\forall f \in \F                       \nn
 & x_f  \leq  \overline{\lambda}_f \,  &&\forall f \in \F  \nn
 & x_f  \geq  \underline{\lambda}_f \, &&\forall f \in \F  \nn
\end{align}
\normalsize
\end{minipage}
\newline
In general, the solution to ${\bf P2}$ need not be the same as the solution to ${\bf
P1}$. However, in our problem, we show that ${\bf P2}$ achieves the solution
of ${\bf P1}$.
\begin{lemma}
\label{lem:opt_joint_vs_sep}.
For a function $f:{\cal Y}\times{\cal Z}\to {\mathbb R}$ that is concave
in $y$ and in $z$, but not jointly in $(y,z)$, the solution to the joint
optimisation problem for convex sets ${\cal Y}$ and ${\cal Z}$ 
\newline
\begin{minipage}{\columnwidth}
\begin{align}
\label{eqn:joint_opt_prob}
\max_{y \in {\cal Y}, z \in {\cal Z}} f(y,z) 
\end{align}
\end{minipage}
is the same as
\newline
\begin{minipage}{\columnwidth}
\begin{align}
\max_{z \in {\cal Z}} \max_{y \in {\cal Y}} f(y,z),\\\nonumber 
\end{align}
\end{minipage}
\newline
if $f(y^*(z),z)$ is a concave function of $z$, where for each $z \in
{\cal Z}$, $y^*(z) := \underset{y \in {\cal
	Y}}{\arg\max} f(y,z)$.
\end{lemma}

We note that for each $x_f$, the probability of error
$e_f(\theta_f,x_f)$ is convex in ${\theta_f}$, and hence, $\ln(1-e_f)$
is concave in $\theta_f$. Thus, we first solve for the optimum Chernoff
bound parameter ${\bm \theta}^*$ which we describe in Section
\ref{subsec:opt_theta}. After having solved for the optimum ${\bm
\theta}^*$, we show in Section~\ref{sec:convexity_condition} that
$U({\bm \theta}^*({\bm x}), {\bm x})$ is a concave function of ${\bm
x}$. Hence, from Lemma~\ref{lem:opt_joint_vs_sep}, the solution to
problem $({\bf P2})$ (the maximisation problem that separately
obtains the optimum ${\bm \theta}^*$ and optimum ${\bm x}^*$) 
is globally optimum. We study the rate optimisation problem that obtains
${\bm x}^*$ in Section \ref{subsec:opt_R}. 

\section{Utility Optimum Rate Allocation} 
\label{sec:sol}

\subsection{Optimal ${\theta}^*$}
\label{subsec:opt_theta}
Consider the following optimisation problem, for any given ${\bm x}
\in [\underline{\lambda}_f,\overline{\lambda}_f]^F$. 
\newline
\begin{minipage}{\columnwidth}
\begin{align}
\max_{\bm \theta} 
& 
\sum_{f \in \F} \ln\left(1-e_f(\theta_f,x_f)\right) \label{eqn:max_theta_problem} \\
\text{subject to} \quad
 &\theta_f  >  0, \ \ \ \ \  \forall f \in \F \nonumber                       
\end{align}
\end{minipage}

\vspace{4mm}

\noindent
We note that the objective function is separable in $\theta_f$s, and
that $e_f$ is convex in $\theta_f$. Hence, the problem defined in
Eqn.~\eqref{eqn:max_theta_problem}, is a concave maximisation problem.
We recall that
\begin{align}
\label{eqn:e1_bound}
e_f(\theta_f,x_f) &= \exp\left(-\frac{k_f}{1-2x_f}\left[
\theta_f x_f - \ln\left(1-\beta_f+\beta_f e^{\theta_f}\right)
\right]\right).
\end{align}
The partial derivative of $e_f$ with respect to $\theta_f$ 
is given by 
\begin{align*}
\frac{\partial e_f}{\partial\theta_f}
&=  e_f \cdot \frac{-k_f}{1-2x_f} \left[ x_f - \frac{\beta_f
e^{\theta_f}}{1-\beta_f+\beta_fe^{\theta_f}}\right]. 
\end{align*}
Observe that $\frac{\beta_f e^{\theta_f}}{1-\beta_f+\beta_fe^{\theta_f}}$ is an
increasing function of $\theta_f$.
Thus, if, for $\theta_f = 0$, $x_f - \frac{\beta_f }{1-\beta_f+\beta_f}
< 0$ or $x_f < \beta_f$ (equivalently, $r_f > 1-2\beta_f$), the
derivative is positive for all $\theta_f >0$, or $e_f$ is an increasing
function of $\theta_f$. Hence, for $x_f < \beta_f$, the optimum
$\theta_f^*$ is arbitrarily close to $0$ which yields $e_f$ 
arbitrarily close to $1$. Thus, for error
recovery, for any end--to--end
error probability $\beta_f$, the coding rate should be smaller than
$1-2\beta_f$, in which case, we
obtain the optimal $\theta_f^*$ by equating 
the partial
derivative of $e_f$ with respect to $\theta_f$ to zero.
\begin{align*}
\begin{array}{lrcl}
\text{i.e.,}&  \frac{\beta_f e^{\theta_f^*}}{1-\beta_f+\beta_fe^{\theta_f^*}} & = & x_f \\
\text{or},  & e^{\theta_f^*} & = & \frac{x_f}{\beta_f}\frac{1-\beta_f}{1-x_f} \\
\text{or},  & \theta_f^* & = & \ln\left(\frac{x_f}{\beta_f}\right) -
 \ln\left(\frac{1-x_f}{1-\beta_f}\right). 
\end{array}
\end{align*}
The probability of error for a given $x_f$ and $\theta_f^*(x_f)$ is then given by
\begin{align}
\label{e_f_star}
& \ \ \ \ e_f(\theta_f^*,x_f)\nn
&=  \exp\left(-\frac{k_f}{1-2x_f}\left[
x_f \ln\left(\frac{x_f}{\beta_f}\right)
+(1-x_f) \ln\left(\frac{1-x_f}{1-\beta_f}\right)
\right]\right)\nn
&= \exp\left(-\frac{k_f}{1-2x_f} D(\mathcal{B}(x_f)||\mathcal{B}(\beta_f)\right)
\end{align}

\subsection{A convex optimisation framework to obtain optimal $x_f^*$}
\label{sec:convexity_condition}
If $\ln(1-e_f(\theta_f^*(x_f),x_f))$ is a concave function of $x_f$, then one can obtain
the optimum $x_f^*$ using convex optimisation framework. 
To show the concavity of $\ln(1-e_f(\theta_f^*(x_f),x_f))$, it is sufficient to
show that $e_f(\theta_f^*(x_f),x_f)$ is convex in $x_f$. Define 
$\Lambda_f := \ln\left(\frac{x_f(1-x_f)}{\beta_f(1-\beta_f)}\right)$. Note that
\begin{align*}
\frac{\partial e_f}{\partial x_f} &= -e_f\cdot\frac{k_f
\Lambda_f}{(1-2x_f)^2}\\
\frac{\partial^2 e_f}{\partial x_f^2} 
 &= \left[
{e_f}\cdot\frac{k_f}{(1-2x_f)^2}
 \right] \\
 &
 \cdot
 \left[
\frac{k_f}{(1-2x_f)^2} \Lambda_f^2
-\frac{4\Lambda_f}{1-2x_f} 
- \frac{1-2x_f}{x_f(1-x_f)}
 \right]
\end{align*}
$e_f(\theta_f^*(x_f),x_f)$ is convex if 
\begin{align*}
\frac{k_f}{(1-2x_f)^2} \Lambda_f^2
&\geq 
\frac{4\Lambda_f}{1-2x_f} 
+ \frac{1-2x_f}{x_f(1-x_f)},
\end{align*}
or, 
\begin{align*}
\frac{4(1-2x_f)}{\Lambda_f} + \frac{(1-2x_f)^3}{x_f(1-x_f)\Lambda_f^2} \leq
 k_f 
\end{align*}
Since, we consider $x_f \geqslant \underline{\lambda}_f$, where 
$\underline{\lambda}_f = \beta_f+\epsilon_f$ for some arbitrarily small 
$\epsilon_f > 0$, we have $\frac{1}{\Lambda_f^2} \leqslant K_0^2$
where $1/K_0 :=
\ln\left(\frac{\underline{\lambda}_f(1-\underline{\lambda}_f)}{\beta_f(1-\beta_f)}\right)$,
and hence, a sufficient condition for the convexity of $e_f$ (and hence,
the concavity of $\ln(1-e_f)$) is 
\begin{align}
\label{new_condition}
\frac{4(1-2x_f)}{\Lambda_f} + K_0^2\frac{(1-2x_f)^3}{x_f(1-x_f)} \leq
 k_f 
\end{align}
The above condition is a convex function of
$x_f$, and we include this as a constraint in the problem formulation.
Thus, $e_f(\theta_f^*(x_f),x_f)$ is convex in $x_f$, and hence, we obtain
the optimal $x_f^*$ using convex optimisation method.
Also, from 
Lemma~\ref{lem:opt_joint_vs_sep}, the optimal coding rate $r_f^* =
1-2x_f^*$ is unique and globally optimum.

\vspace{4mm}
The minimum $k_f$ required to ensure convexity of
$e_f(\theta_f^*(x_f),x_f)$ is computed numerically, and is tabulated
below. 

\vspace{2mm}
\begin{table}[h]
\begin{center}
\caption{Minimum $k_f$ that ensures convexity of
$e_f(\theta_f^*(x_f),x_f)$}
\begin{tabular}{|l|c|}
\hline
$\beta_f$ & minimum $k_f$ required\\
\hline
0.1       & 6  \\
0.01      & 10 \\
0.001     & 33 \\
0.0001    & 164\\
\hline
\end{tabular}
\end{center}
\end{table}

From the above table, we see that the minimum packet size required to
ensure convexity is very small, and in practice, the packet size $k_f$
is much larger than the minimum size required. Hence, for all practical
purposes, the optimal code rate problem is a convex problem. More
importantly, the constraint given by Eqn.~\eqref{new_condition} is
not an active constraint. However, for the sake of completeness, we include this
constraint in the problem definition below. 

\subsection{Optimal Coding Rate $\boldmath{r}$}
\label{subsec:opt_R}
In this subsection, we obtain the optimal coding rate using the optimal
Chernoff--bound parameter vector ${\bm
\theta}^*$, by solving the following network utility maximisation problem 
\newline
\begin{minipage}{\columnwidth}
\begin{align}
\max_{\bm x} 
& 
\sum_{f \in \F} \ln\left(1-e_f(\theta_f^*,x_f)\right) \label{eqn:max_x_problem} \\
\text{subject to} \quad
&\underset{f: c \in \C_f  }{\sum} \frac{k_f}{(1-2x_f)w_{f,c}}  \leq  T_c,  &&\forall c \in \C \nonumber \\ 
 & x_f  \leq  \overline{\lambda}_f \,  &&\forall f \in \F  \nn
 & x_f  \geq  \underline{\lambda}_f \, &&\forall f \in \F  \nn
& \frac{4(1-2x_f)}{\Lambda_f} + K_0^2\frac{(1-2x_f)^3}{x_f(1-x_f)} \leq
 k_f \, &&\forall f \in \F  
 \label{eqn:max_x_problem_k_constraint} 
\end{align}
\end{minipage}
\newline
The objective function is separable and concave, and
hence, can be solved using Lagrangian relaxation method. Also, 
the constraint represented
by Eqn.~\eqref{eqn:max_x_problem_k_constraint} is not an active constraint,
and hence, there is no Lagrangian cost to this constraint. We note here
that the coding rate should be such that $k_f/(1-2x_f)$ is an integer,
and hence, obtaining $x_f^*$ is a discrete
optimisation problem. This is, in general, an NP hard problem. Hence, we relax
this constraint, and allow $x_f$ to take any real value in  
$[\underline{\lambda}_f, \overline{\lambda}_f]$. 
The
Lagrangian function for the optimal rate problem is thus
\begin{align*}
& \ \ \ \ L({\bm x},{\bm p},
{\bm u},
{\bm v} 
)\\
&= 
\sum_{f \in \F} \ln\left(1 - e_f(\theta_f^*,x_f)\right)
- \sum_{c \in \C}p_c\left(\sum_{f \in \F_c} \frac{k_f}{(1-2x_f)w_{f,c}}
  - T_c\right)\\
  &
+ \sum_{f \in \F}u_f\left(x_f - \underline{\lambda}_f\right) 
- \sum_{f \in \F}v_f\left(x_f - \overline{\lambda}_f\right) 
\end{align*}
Applying KKT condition,
$\frac{\partial L}{\partial x_f}\mid_{x_f^*} = 0$,
we have
\begin{align*}
\frac{-1}{1-e_f} \frac{\partial e_f}{\partial x_f}\mid_{x_f^*} 
&= \sum_{c \in \C_f}\frac{p_c}{w_{f,c}} \frac{2k_f}{(1-2x_f^*)^2} + v_f - u_f\\ 
&= 
\frac{2k_f}{(1-2x_f^*)^2} \left( \sum_{c \in \C_f}\frac{p_c}{w_{f,c}}
\right) + v_f - u_f\\ 
%
 \frac{e_f}{1-e_f}\cdot\frac{k_f\Lambda_f^*}{(1-2x_f^*)^2}
&=
\frac{2k_f}{(1-2x_f^*)^2} \left( \sum_{c \in \C_f}\frac{p_c}{w_{f,c}}
\right) + v_f - u_f\\ 
 \frac{e_f}{1-e_f}\Lambda_f^*
&= 2 \left( \sum_{c \in \C_f}\frac{p_c}{w_{f,c}}  \right) +
\frac{(v_f-u_f)(1-2x_f^*)^2}{k_f}\\
&= \lambda_f + 
\frac{(v_f-u_f)(1-2x_f^*)^2}{k_f}
\end{align*}
where  $\lambda_f :=   2 \left( \sum_{c \in \C_f}\frac{p_c}{w_{f,c}}
\right)$ and $\Lambda_f^* :=
\ln\left(\frac{x_f^*(1-x_f^*)}{\beta_f(1-\beta_f)}\right)$.
If the optimal $x_f^*$ is either $\underline{\lambda}_f$ or 
$\overline{\lambda}_f$, then it is unique. If 
$x_f^* \in (\underline{\lambda}_f, \overline{\lambda}_f)$, then $u_f
= v_f = 0$, and in this case 
(which is the most
interesting case, and we consider only this case for the rest of the paper), 
we have  
\begin{align}
\frac{e_f}{1-e_f} \cdot \Lambda_f^* 
&= \lambda_f \nn
e_f &=
							  \frac{\lambda_f}{\lambda_f+\Lambda_f^*
							  }\label{eqn:ef}\\ 
\exp\left(-\frac{k_f}{1-2x_f^*} D({\cal B}(x^*_f)\|{\cal B}(\beta_f))\right)
&=
                              \frac{\lambda_f}{\lambda_f+\Lambda_f^*
							  }\nn 
\frac{k_f}{1-2x_f^*} D({\cal B}(x_f^*)\|{\cal B}(\beta_f))
&=
\ln\left(\frac{\lambda_f+\Lambda_f^*}{\lambda_f}\right)
\label{eqn:x_star}
\end{align}
In the above equation, both the LHS and the RHS are increasing in $x_f^*$.
Also, LHS is a strictly convex (increasing) function and RHS is a
strictly concave (increasing) function of $x_f^*$. Hence, they intersect
at exactly one point in the region $(\beta_f, 0.5]$ which is the optimal
$x_f^*$ for a given Lagrangian price vector ${\bm p}$.

\subsection{Sub--gradient Approach to Compute optimal ${p}_c^*$}
In this section, we discuss the procedure to obtain the optimal shadow
costs or the Lagrange variables ${\bm p}^*$. The dual problem for the primal
problem defined in Eqn.~\eqref{eqn:max_x_problem} is given by
\begin{eqnarray*}
\min_{{\bm p} \geq 0} & & D(\bm{p}),
\end{eqnarray*}
where the dual function $D(\bm{p})$ is given by
\footnotesize
\begin{align}
& \hspace{5mm} D(\bm{p})\nn
&= \max_{{\bm x}} \underset{f\in\F}{\sum}\ln(1-e_f(x_f))+\underset{c\in\C}{\sum}p_c\left( T_c - \underset{f \in \F_c}{\sum}
\frac{k_f}{(1-2x_f)w_{f,c}}
\right) \label{eq:supremum}\\
&= \underset{f\in\F}{\sum}\ln(1-e_f(x_f^*({\bm
p})))+\underset{c\in\C}{\sum}p_c\left( T_c - \underset{f \in \F_c}{\sum}
\frac{k_f}{(1-2x_f^*({\bm p}))w_{f,c}} 
\right).\\\nonumber
 \label{eq:d_star}
\end{align}
\normalsize

\noindent
In the above equation, $e_f(x_f)$ denotes $e_f(\theta_f^*(x_f),x_f)$.
Since the dual function (of a primal problem) is
convex, $D$ is convex in ${\bm p}$. Hence, we use a sub--gradient method
to obtain the optimum $\bm{p}^*$.
From Eqn.~\eqref{eq:supremum}, for any ${\bm x}$, 

\footnotesize
\begin{align*}
D(\bm{p}) &\geq \underset{f\in\F}{\sum}\ln(1-e_f(x_f))+\underset{c\in\C}{\sum}p_c\left( T_c - \underset{f \in \F_c}{\sum}
\frac{k_f}{(1-2x_f)w_{f,c}}
\right),\\
\end{align*}
\normalsize
and in particular, the dual function  $D({\bm p})$ is greater than that
for $x=x_f^*({\widetilde{\bm p}})$, i.e.,
\footnotesize
\begin{align}
& \hspace{5mm}D(\bm{p})\nn
&\geq \underset{f\in\F}{\sum}\ln(1-e_f(x_f^*({\widetilde{\bm p}})))+\underset{c\in\C}{\sum}p_c\left( T_c - \underset{f \in \F_c}{\sum}
\frac{k_f}{(1-2x_f^*({\widetilde{\bm p}}))w_{f,c}}
\right)\nn
&= D(\widetilde{\bm p})
+\underset{c\in\C}{\sum}\left(p_c-\widetilde{p}_c\right)\left( T_c - \underset{f \in \F_c}{\sum}
\frac{k_f}{(1-2x_f^*({\widetilde{\bm p}}))w_{f,c}}
\right)
\end{align}
\normalsize
Thus, a sub--gradient of $D(\cdot)$ at any $\widetilde{\bm p}$ is given
by the vector 
\begin{align}
\left[ T_c - \underset{f \in \F_c}{\sum}
\frac{k_f}{(1-2x_f^*({\widetilde{\bm p}}))w_{f,c}}\right]_{c \in
\C}.\\\nonumber
\end{align}
We obtain an iterative algorithm based on sub--gradient method that
yields $\bm{p}^*$, with ${\bm p}(i)$ being the Lagrangians at the $i$th
iteration. 
\begin{eqnarray*}
p_c(i+1) = \left[p_c(i)-\gamma\cdot
\left(T_c - \underset{f \in \F_c}{\sum}
\frac{k_f}{(1-2x_f^*({\bm p}(i)))w_{f,c}} 
\right)\right]^+
\end{eqnarray*}
where $\gamma > 0$ is a sufficiently small stepsize, and $[f(x)]^+ := \max\{f(x),0\}$
ensures that the Lagrange multiplier
never goes negative. Note that the Lagrangian updates can be locally
done, as each cell $c$ is required to know only the rates $x_f^*({\bm
p}(i))$ of flows $f \in \F_c$. Thus, at the beginning of each iteration
$i$, the flows choose their coding rates to $1-2x_f^*({\bm p}(i))$, and each cell
computes its cost based on the rates of flows through it. The updated
costs along the route of each flow are then fed back to the source node
to compute the rate for the next iteration.

The Lagrange multiplier $p_c$ can be viewed as the cost of transmitting
traffic through cell $c$. The amount of service time that is available
is given by $\Delta = T_c - \underset{f \in \F_c}{\sum}
\frac{k_f}{(1-2x_f^*({\bm p}(i)))w_{f,c}}$. When $\Delta$ is positive and
large, then the Lagrangian cost $p_c$ decreases rapidly (because $D$ is
convex), and when $\Delta$ is negative, then the Lagrangian cost $p_c$
increases rapidly to make $\Delta \geq 0$. We note that the increase or
decrease of $p_c$ between successive iterations is proportional to
$\Delta$, the amount of service time available. Thus, the sub--gradient
procedure provides a dynamic control scheme to balance the network load.

We explore the properties of the optimum rate parameter $x_f^*$ in 
Section~\ref{subsec:x_f_s}. In Section~\ref{sec:examples}, we provide
some examples that illustrate the optimum utility--fair resource
allocation.

\subsection{Properties of $x_f^*$}
\label{subsec:x_f_s}

We are interested in studying the behaviour of the optimum coding
rate $r_f^* = 1-2x_f^*$, when the PHY rate $w_{f,c}$ and the packet size
$k_f$ increases such that $k_f/w_{f,c}$ is always a constant.
\begin{lemma}
\label{lem:monotone_x_star}
$r_f^* = 1-2x_f^*(k_f)$ is an increasing function of $k_f$ (with the PHY
rate $w_{f,c}$ being proportional to $k_f$).
\end{lemma}

Lemma~\ref{lem:monotone_x_star} is quite intuitive. For any given
channel error $\beta_f$, as the block (or packet) length increases, it
is optimum to go for a high rate code. In other words, it is optimum for
a flow to use as much scheduling time as possible (i.e., use a large
block length $k_f$, and hence, use a high rate code); however, the resources are shared among
multiple flows, and hence, we ask the following question:  ``{\em what is
the optimum share of the scheduling time}'' that each flow should have.
Interestingly, in our problem formulation, the optimum code rate
parameter $x_f^*$ also solves this optimum scheduling times for each
flows. 

It is interesting to ask the question of {\em how large the
packet sizes $k_f$ be for optimum resource allocation,} and
Lemma~\ref{lem:monotone_x_star} provides a hint to the solution.
From Lemma~\ref{lem:monotone_x_star}, we understand the
following: if there are two flows $f_1, f_2$, through a
cell $c$ (seeing the same channel conditions, i.e., $\beta_{f_1} =
\beta_{f_2}$) with $w_{f_1,c} > w_{f_2,c}$ then it is optimum for flow $f_1$ to
use a large packet size $k_{f_1}$ and flow $f_2$ to
use a small packet size $k_{f_2}$. The optimum schedule length will be
to allocate less schedule time
to flow $f_1$ and more schedule time to flow $f_2$. 

\vspace{5mm}

\noindent
In the {\em asymptotic case when $w_{f,c}$ and $k_f$ grows to $\infty$
(and $k_{f}$ grows linearly with $w_{f,c}$}, we see from
Eqn.~\eqref{eqn:x_star} that the error exponent also goes to $\infty$ 
(as $1-2x_f > 0$), and hence, $e_f \to 0$. In this case, we see that the
optimum rate can approach arbitrarily close to $1-2\beta_f^*$. Thus, for 
any $k_f$ and $w_{f,c}$, the optimum coding rate 
$r_f^* < 1-2\beta_f^*$

\vspace{4mm}

Previous studies on optimum
resource allocation   
establish that the proportional fair allocation is the same as 
equal air--time allocation (\cite{eq_air_time}). 
But, in this problem, we see an interesting phenomenon that is unusual of a
proportional--fair resource allocation. 
\begin{lemma}
\label{lem:unequal_air_time}
The optimum rate allocation ${\bm x}^*$ (or equivalently ${\bm r}^*$) is
not equivalent to equal air--time allocation which is typically the
solution of a proportional--fair (or $\ln$ utility) allocation. 
\end{lemma}

In particular, we see that the flows that see a better channel get less
air--times than the flows that see a worse channel. This phenomenon is
evident in the case of infinitely long code words; with other parameters
being same, the air--times of flows in a cell $c$ are proportional to 
$\frac{1}{1-2\beta_{f,c}}$, and hence, flows with small $\beta$ get
less air--times. 

\section{Examples}
\label{sec:examples}

In this Section, we analyse some simple networks based on the utility
optimum solution that we obtained. In particular, we analyse the 
so--called parking--lot topology often used to explore fairness issues.
It is to be noted that the parking--lot topology is a simple case of a
line network, and the results of this section extends in a simple way to
a linear network.

\begin{figure}[t]
\begin{center}
\includegraphics[width=50mm, height=35mm]{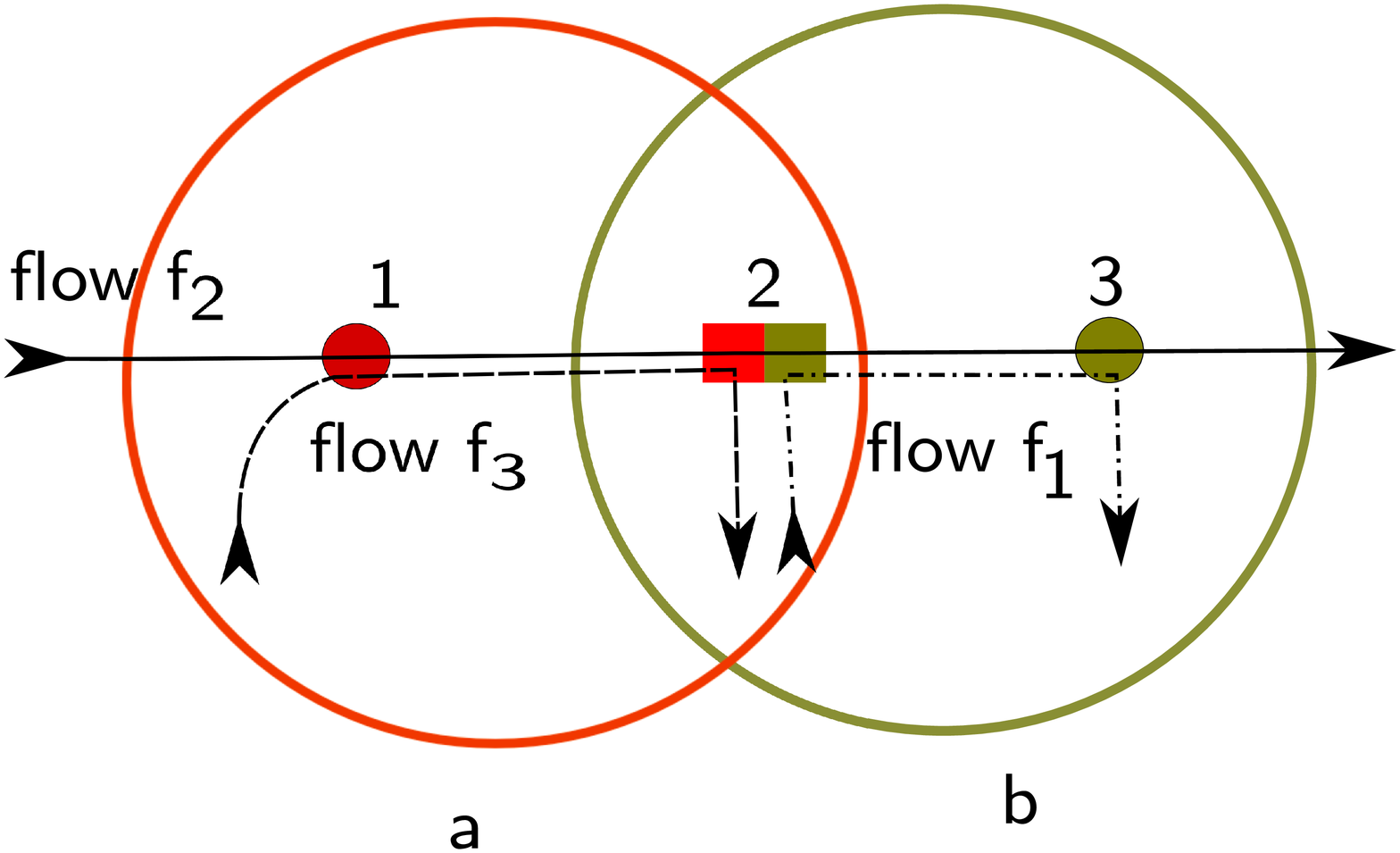}
\caption{Cells with equal traffic load}
\label{fig:example_equal_traffic}
\end{center}
\end{figure}
\subsection{Example 1: Two cells with equal traffic load}

We begin by considering the example shown in
Figure~\ref{fig:example_equal_traffic} consisting of two cells $a$ and
$b$ having three nodes 1, 2, and 3. Each cell has the same symbol 
error probability $\beta$ and the schedule length $T$.  There
are three flows $f_1, f_2$, and $f_3$, with two of the flows $f_1$ and
$f_3$ having one--hop routes $\C_{f_1} = \{b\}$ and $\C_{f_3} = \{a\}$,
and one flow $f_2$ having a two--hop route $\C_{f_2} = \{a,b\}$. Each
flow has the same information packet size $k$ and
PHY transmit rate, \emph{i.e.} $w_{f,c} = w$. 

The end--to--end packet error probability experienced by the two--hop
flow $f_2$ is greater than that experienced by the one hop flows $f_1$
and $f_3$, since each hop has the same fixed error probability. Hence,
we need to assign a lower coding rate $r_{f_2}$ to flow $f_2$ than to
flows $f_1$ and $f_3$ in order to obtain the same error probability
(after decoding) across flows. However, when operating at the boundary
of the network capacity region (thereby maximising throughput),
decreasing the coding rate $r_{f_2}$ of the two--hop flow $f_2$ requires
that the coding rate of \emph{both} one--hop flows  $f_1$ and $f_3$ be
increased in order to remain within the available network capacity. In
this sense, allocating coding rate to the two--hop flow $f_2$ imposes a
greater marginal cost on the network (in terms of the sum--utility) than
the one--hop flows, and we expect that a fair allocation will therefore
assign higher coding rate to the two--hop flow $f_2$. The solution
optimising this trade--off in a proportional fair manner can be
understood using the analysis in the previous section. 

In this example, both the cells are equally loaded and, by symmetry, the
Lagrange multipliers $p_a = p_b$. Hence, $\lambda_{f_1} =
\frac{\lambda_{f_2}}{2} = \lambda_{f_3}$. Note that $x_{f_2}^* <
x_{f_1}^*$ and $\Lambda_{f_2}^* < \Lambda_{f_1}^*$.
Hence, we find from 
Eqn.~\eqref{eqn:ef} that
\begin{eqnarray*}
\frac{e_{f_1}}{e_{f_2}} & = &  \frac{\lambda_{f_1}}{\lambda_{f_2}}
\frac{\lambda_{f_2}+\Lambda_{f_2}^*}{\lambda_{f_1}+\Lambda_{f_1}^*}\nn
&<& 1.
\end{eqnarray*}

\begin{figure}[t]
\centering
\includegraphics[width=50mm, height=35mm]{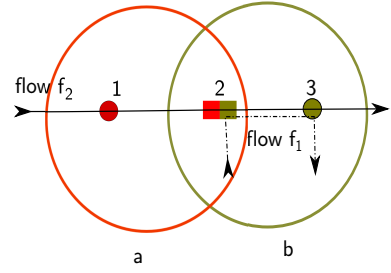}
\caption{Cells with unequal traffic load}
\label{fig:example_unequal_traffic}
\end{figure}


\subsection{Example 2: Two cells with unequal traffic load}

We consider the same network as in the previous example, but now with
only the flows $f_1$ and $f_2$ (i.e., the flow $f_3$ is not present, 
see Figure~\ref{fig:example_unequal_traffic}) in
the network. In this example, cell $b$ carries two flows while cell $a$
carries only one flow. The encoding rate constraints are given by
\begin{eqnarray*}
\frac{1}{r_{f_2}}  & \leq & \frac{wT}{k}, \ \text{(from cell $a$)}, \\ 
\frac{1}{r_{f_1}} + \frac{1}{r_{f_2}} & \leq & \frac{wT}{k}, \ \text{(from cell $b$)}.
\end{eqnarray*}
Since, both $r_{f_1}$ and $r_{f_2}$ are at most 1, it is clear that at
the optimum point, the rate constraint of cell $a$ is not tight while
the constraint of cell $b$ is tight. Thus, the shadow prices (Lagrange
multipliers) $p_a = 0$ and $p_b >0$.   That is, at the first hop the
cell is not operating at capacity, and so the ``price'' for using this
cell is zero.  In this example, $\lambda_{f_1} = \lambda_{f_2}$, and
hence, from Eqn.~\eqref{eqn:ef}, we deduce that for low channel errors,
$e_{f_1} \approx e_{f_2}$. This
allocation make sense intuitively since although flow $f_2$ crosses two
hops, it is only constrained at the second hop and so it is natural to
share the available capacity of this second hop approximately equally
between the flows. 

\section{Conclusions}
In this paper, we posed a utility fair problem that yields the optimum
coding across flows in a capacity constrained network. We showed that
the problem is highly non--convex. However, we provided some simple
conditions under which the global network utility optimisation problem
can be solved. 
We obtained the optimum coding rate, and analysed some of its
properties. We also analysed some simple networks
based on the utility optimum framework we proposed. To the best of our
knowledge, this is the first work on cross--layer optimisation that
studies optimum coding across
flows which are competing for network resources.

\bibliographystyle{IEEEtran}
\bibliography{IEEEabrv,allerton}

\begin{thebibliography}{1}
\providecommand{\url}[1]{#1}
\csname url@samestyle\endcsname
\providecommand{\newblock}{\relax}
\providecommand{\bibinfo}[2]{#2}
\providecommand{\BIBentrySTDinterwordspacing}{\spaceskip=0pt\relax}
\providecommand{\BIBentryALTinterwordstretchfactor}{4}
\providecommand{\BIBentryALTinterwordspacing}{\spaceskip=\fontdimen2\font plus
\BIBentryALTinterwordstretchfactor\fontdimen3\font minus
  \fontdimen4\font\relax}
\providecommand{\BIBforeignlanguage}[2]{{%
\expandafter\ifx\csname l@#1\endcsname\relax
\typeout{** WARNING: IEEEtran.bst: No hyphenation pattern has been}%
\typeout{** loaded for the language `#1'. Using the pattern for}%
\typeout{** the default language instead.}%
\else
\language=\csname l@#1\endcsname
\fi
#2}}
\providecommand{\BIBdecl}{\relax}
\BIBdecl

\bibitem{cover_book}
T.~M. Cover and J.~A. Thomas, \emph{Elements of information theory},
  1st~ed.\hskip 1em plus 0.5em minus 0.4em\relax New York: Wiley--Interscience,
  1991.

\bibitem{mushkin1989capacity}
M.~Mushkin and I.~Bar-David, ``Capacity and coding for the {G}ilbert--{E}lliot
  channels,'' \emph{Information Theory, IEEE Transactions on}, vol.~35, no.~6,
  pp. 1277--1290, 1989.

\bibitem{erasure}
K.~Premkumar, X.~Chen, and D.~J. Leith, ``Utility optimal coding for packet
  transmission over wireless networks -- {Part II}: Networks of packet erasure
  channels,'' in \emph{submitted}, 2011.

\bibitem{eq_air_time}
A.~Checco and D.~J. Leith, ``Proportional fairness in 802.11 wireless lans,''
  \emph{to appear in IEEE Comm. Letters}, 2011.

\bibitem{net-opt}
S.~Shakkottai and R.~Srikant, \emph{Network Optimization and Control}.\hskip
  1em plus 0.5em minus 0.4em\relax Now Publishers Inc., Boston - Delft, 2008.

\bibitem{mc_williams_sloane}
F.~J. MacWilliams and N.~J.~A. Sloane, \emph{The theory of error-correcting
  codes}.\hskip 1em plus 0.5em minus 0.4em\relax North-Holland Publishing Co.,
  Amsderdam, 1977.

\bibitem{mds_conv_codes}
R.~Smarandache, H.~Gluesing-Luerssen, and J.~Rosenthal, ``Constructions of
  mds-convolutional codes,'' \emph{Information Theory, IEEE Transactions on},
  vol.~47, no.~5, pp. 2045 --2049, jul 2001.

\end{thebibliography}
\end{document}